\documentclass[sigconf]{acmart}

\AtBeginDocument{%
  }

\usepackage{multirow}
\usepackage{xcolor}
\usepackage{algorithm}
\usepackage{algorithmic}

\setcopyright{acmlicensed}
\copyrightyear{2024}
\acmYear{2024}
\acmDOI{XXXXXXX.XXXXXXX}

\acmConference[MFFM Workshop @ ICAIF '24]{International Workshop on Multimodal Financial Foundation Models (MFFMs) at 5th ACM International Conference on AI in Finance}{November 15, 2024}{Brooklyn, NY,}
\acmISBN{978-1-4503-XXXX-X/18/06}

\begin{document}

\title{\textsc{RiskLabs}: Predicting Financial \underline{Risk} Using \underline{La}rge Language Model \underline{b}ased on Multimodal and Multi-Source\underline{s} Data}

\author{%
  Yupeng Cao$^{+}$, 
  Zhi Chen, 
  Prashant Kumar,
  Qingyun Pei,
  Yangyang Yu,
  Haohang Li,\\
  Fabrizio Dimino,
  Lorenzo Ausiello,
  K.P. Subbalakshmi$^{+}$,
  Papa Momar Ndiaye
}

\affiliation{%
  \institution{Stevens Institute of Technology
  }
  \country{}
  \city{}
}
\email{ycao33@stevens.edu, ksubbala@stevens.edu}

\renewcommand{\shortauthors}{Cao et al.}
\renewcommand{\shorttitle}{RiskLabs}

\begin{abstract}
 The integration of Artificial Intelligence (AI) techniques, particularly large language models (LLMs), in finance has garnered increasing academic attention. Despite progress, existing studies predominantly focus on tasks like financial text summarization, question-answering (Q$\&$A), and stock movement prediction (binary classification), the application of LLMs to financial risk prediction remains underexplored. Addressing this gap, in this paper, we introduce \textbf{RiskLabs}, a novel framework that leverages LLMs to analyze and predict financial risks. RiskLabs uniquely integrates multimodal financial data—including textual and vocal information from Earnings Conference Calls (ECCs), market-related time series data, and contextual news data—to improve financial risk prediction. Empirical results demonstrate RiskLabs' effectiveness in forecasting both market volatility and variance. Through comparative experiments, we examine the contributions of different data sources to financial risk assessment and highlight the crucial role of LLMs in this process. We also discuss the challenges associated with using LLMs for financial risk prediction and explore the potential of combining them with multimodal data for this purpose. \footnote{Position Paper}\footnote{Present at International Workshop on Multimodal Financial Foundation Models (MFFMs) @ ICAIF'24}\footnote{\url{https://sites.google.com/view/iwmffm2024/accepted-papers?authuser=1}}\footnote{ This position work was completed in April 2024.}
\end{abstract}




\maketitle

\section{Introduction}
The integration of artificial intelligence (AI) technology for financial risk prediction has been a longstanding focus of both industry and academia. Previously, researchers used machine learning techniques to analyze historical stock price data for risk prediction, such as using SVM to predict credit ratings based on financial statement ratios~\cite{khaidem2016predicting, lee2007application} and using tree-based models to explore the correlation between economic indicators and the performance of various asset classes~\cite{gu2020empirical}. Recently, researchers have increasingly extracted trading signals from publicly available information for risk prediction. For example, financial reports and media news are used to analyze sentiment in news articles, capture market expectations for stock movements~\cite{souma2019enhanced, mohan2019stock, liapis2023investigating}, risk factor analysis~\cite{zhang2020financial}, and identify early warning signals from large amounts of textual data~\cite{ahbali2022identifying, wang2023sparsity}. With the rise of multimodal learning, more unstructured multimedia data, such as audio recordings of earning conference calls~\cite{qin2019you, yang2020html} are utilized for stock volatility prediction.

While supervised learning methods have proven effective, they are highly task-specific and lack adaptability~\cite{singh2016review}, with their predictive performance constrained by the amount of input data and the size of parameters they can handle. However, the emergence of large language models (LLMs) represents a paradigm shift in overcoming these challenges. Equipped with vast knowledge bases and advanced zero-shot learning capabilities, LLMs can perform a wide range of text-related tasks, such as summarization~\cite{zhang2024benchmarking}, question answering~\cite{wei2022chain}, and sentiment analysis~\cite{zhang2023enhancing}, without requiring task-specific training. In the financial domain, LLMs have been widely utilized for tasks such as financial text analysis~\cite{yang2023fingpt} and financial report generation~\cite{xie2024finben}. Furthermore, LLM-based agent systems have been employed for stock trading~\cite{yu2023finmem, zhang2024ai, zhang2024finagent}. These studies highlight the versatility and impact of LLMs in finance, emphasizing their growing significance in enhancing business operations and decision-making processes.

\begin{figure*}[h]
    \centering
    \includegraphics[width=0.85\textwidth]{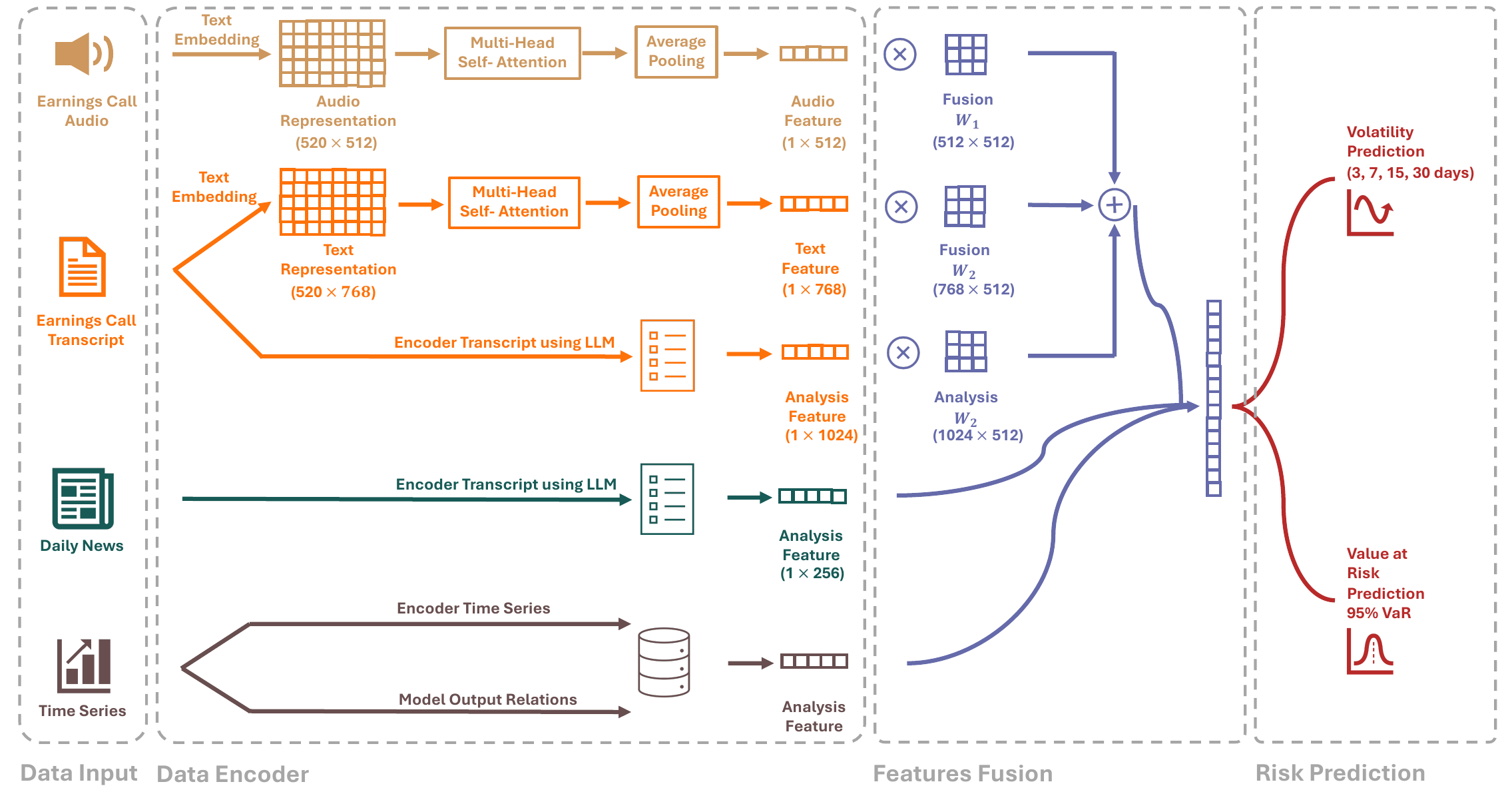}
    \caption{This figure illustrates the RiskLabs Framework. The model accepts multiple-source inputs: Earnings Conference Call Audio \& Transcript, Daily News, and Time Series Data. The second area visualizes the model's pipeline to encode diverse sources and illustrates how LLMs are applied for data analysis. The third area describes how the model consolidates outputs from both embeddings and LLM analysis for use in subsequent stages. The model will perform multi-task learning: our RiskLabs will predict the Volatility of different terms and VaR in the meantime.}
    \label{Framework}
\end{figure*}

Despite the expanding body of work on LLM applications in finance, there remains a significant gap in exploring their potential for financial risk prediction, particularly for metrics like stock volatility and Value at Risk (VaR). In addition, since LLMs are generative models, the challenge, of using them for precise numerical regression~\cite{song2024omnipred}, also persists in financial risk prediction tasks. Therefore, our research aims to fill this gap by leveraging LLMs for financial risk prediction, contributing valuable insights to understanding market dynamics and ensuring sustainable success across various sectors. This position paper is interested in exploring the following research questions \textbf{(RQs)}:
\begin{itemize}
    \item \textbf{RQ1}: What role do large language models play in financial risk prediction?
    \item \textbf{RQ2}: How does the predictive performance of Large Language Models (LLMs) compare with other artificial intelligence techniques in forecasting risk metrics?
    \item \textbf{RQ3}: How can multiple inputs of varying data types be effectively incorporated and balanced?
\end{itemize}

To address the issues previously mentioned, we propose the \textsc{RiskLabs} framework, designed to comprehensively understand the complexities of the investment environment. This framework ingests and analyzes multimodal data from multiple sources, providing a holistic view of the factors influencing financial markets. The multimodal data sources include: (1) earnings conference call transcripts, offering insights into corporate performance and future prospects; (2) audio from these calls, capturing tone and sentiment not evident in the text; (3) time series data for historical and real-time market trend analysis and (4) media news. Leveraging LLMs, \textsc{RiskLabs} extracts essential information from each modality and applies multimodal fusion techniques to integrate these features, enhancing financial risk prediction.

RiskLabs is equipped with four key modules to effectively process these diverse data streams: a) The Earnings Conference Call Encoder: leveraging LLM to handle data related to earnings conference call (1 and 2); b) The News-Market Reactions Encoder: establishing a pipeline through LLM to collect and interpret news data; c) The Time-Series Encoder: organizing and analyzing time-related data; and d) The Multi-Task Prediction: which amalgamates outputs from the aforementioned modules for multifaceted prediction. This synthesis of varied inputs allows the model to provide a nuanced understanding of the investment landscape, blending quantitative market data with qualitative insights. Such an integrated methodology is crucial for precise forecasting and risk assessment, a key to navigating the often turbulent and complex terrain of financial markets. By incorporating a diverse range of inputs and using processing steps, RiskLabs solves a multitasking problem: forecasting volatility over various intervals (3 days, 7 days, 15 days, and 30 days), as well as the value at risk (VaR).

Our contributions are encapsulated as follows: 1) We utilize LLM to develop the \textsc{RiskLabs} framework, thereby addressing the shortfall in applying LLMs to financial risk prediction. 2) The versatility of our model lies in its capability to seamlessly integrate a variety of financial multimodal data from multiple sources, consequently bolstering risk prediction accuracy. 3) The efficacy of our framework is evidenced by experimental results, demonstrating its effectiveness in forecasting financial risks. 4) Furthermore, we provide a comprehensive analysis of the pivotal role LLM should assume in the realm of financial risk prediction.  

\section{\textsc{RiskLabs} Framework}\label{s2}
Figure \ref{Framework} illustrates the RiskLabs Framework, designed to handle multiple data types surrounding the financial information including audio, text, and time-series from different sources. The framework comprises four main modules: 1) Earnings Conference Call Encoder; 2) Time-Series Encoder; 3) Relevant News Encoder; and 4) Multimodal Fusion and Multi-Task Prediction. This section outlines the format of the data input and details each module. In brief, the Earnings Conference Call Encoder, Time-Series Data Encoder, and Relevant News Encoder are utilized to extract features from various data types. These features are fused data that undergoes processing and modeling, after which it is fed into the Multi-Task Prediction Block. This block is responsible for forecasting both volatility for different intervals and VaR (Value at Risk) values.

\begin{figure*}[t]
\centering
  \includegraphics[width=0.80\linewidth]{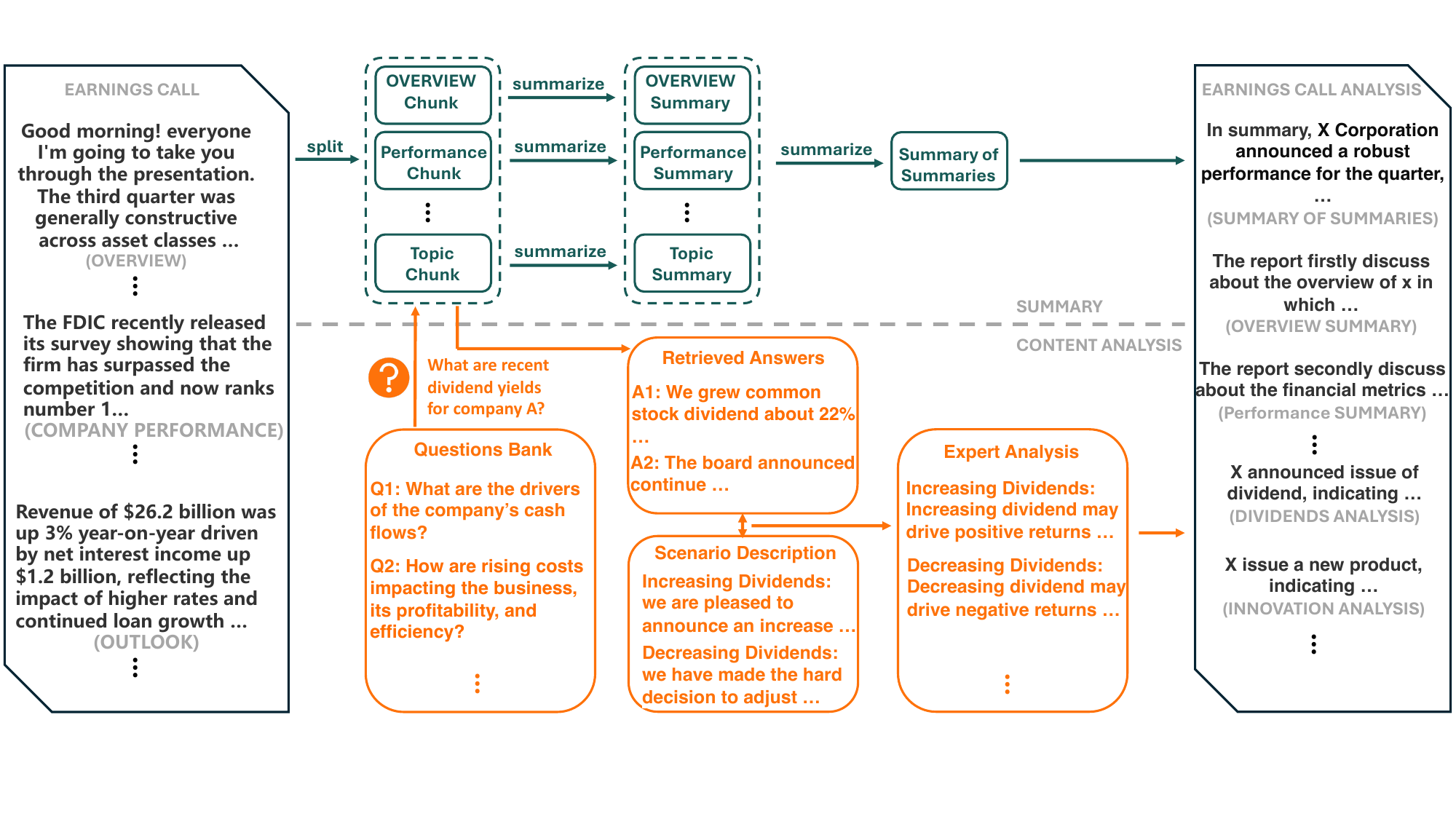} 
  \caption {visualizes the process of fine-grained information extraction from ECC transcript. 
  }
  \label{ECC_Transcript_Analysis}
\end{figure*}

\subsection{Earnings Conference Call Encoder} 
The Earnings Conference Call Encoder module consists of three main components: Audio Encoding, Transcript Encoding, and Earnings Conference Call Analyzer.

\subsubsection{Audio Encoding} Audio embeddings are extracted using Wav2vec2 \cite{baevski2020wav2vec} and then processed through a Multi-Head Self-Attention (MHSA) layer with pooling to obtain audio features $T_a$. To describe the Audio Encoding in more detail, we let the raw audio input data be represented by \(A_c = \{a_{c}^1, a_{c}^2, \ldots, a_{c}^n \}\) where \(a_c^i\) represents the \(i^{th}\) audio frame in one data sample. Each audio frame will be converted into a vector representation:
\begin{equation}
e_{ac}^i = \text{Wav2Vec2}(a_c^i).
\end{equation}

 We obtain the audio embeddings \(E_{ac} = \{e_{ac}^1, e_{ac}^2, \ldots, e_{ac}^n \}\). The shape of the \(E_{ac}\) is 520 $\times$ 512, where 520 is the maximum number of audio files amongst all companies and 512 is the dimension of the resulting transform from the model for a single audio frame. Earnings conference calls with less than 520 audio frames ($n < 520$) have been zero-padded for uniformity in input matrix size. 

Then, we fed audio embedding \(E_{ac}\) into the Multi-Head Self-Attention (MHSA) to extract the audio feature. The MHSA further comprises a multi-head attention block, followed by a norm block, and an MLP block. MLP denotes a two-layer feed-forward network (FNN) with a ReLU activation function. The MHSA block essentially forms the basis of all the architectures discussed later in the paper. In detail, the multi-head self-attention calculation process is as follows:
\begin{equation}
\text{Multihead} =  \text{Concat}\text{(head}_1, ...., \text{head}_h)W^o
\end{equation}
\begin{equation}
\text{head}_i =  \text{Attention}(QW_i^Q, KW_i^K, VW_i^V)
\end{equation}

where $Q$ (queries) and $K$ (keys) of dimension $d_k$ and $V$ values of dimension $d_v$. The dimensions of the weights are: $W_i^Q \in R^{d_{model} \times d_k}$, $W_i^K \in R^{d_{model}\times d_k}$, $W_i^V \in R^{d_{model} \times d_v}$, and $W^o \in R^{d_{v}\times d_{model}}$. The dot product is then calculated for the query with all the keys. These values are then normalized by dividing each value by $\sqrt{d_k}$ and then we apply a softmax function to obtain the weights of the values:
\begin{equation}
\text{Attention}(Q, K, V) = softmax(\frac{KQ^T}{\sqrt{d_k}})V
\end{equation}

The attention function on a set of queries is calculated simultaneously packed together in a matrix Q. The keys and values are also packed in the matrices K and V respectively. 

Combining (2)-(4), we apply this process to \(E_{ac}\):
\begin{equation}
T_{ac} =  \text{MHSA}(E_{ac})
\end{equation}
where \(T_{ac} = \{t_{ac}^1, t_{ac}^2, \ldots, t_{ac}^n \}\) with size 520 $\times$ 512. Following this, an average pooling layer is applied to \(T_{ac} \):
\begin{equation}
T_{a} =  \text{AveragePooling}(T_{ac})
\end{equation}
where  \(T_{a}\) denotes the resultant extracted audio feature of size 512.

\subsubsection{Transcript Encoding.} Similar to Audio Encoding, this component first uses SimCSE \cite{gao2021simcse} to generate vector representations for each sentence in the earnings call transcripts. These embeddings are then passed through an MHSA layer with pooling to capture text features $T_t$; We let raw transcripts as \(T_c\ = \{t_{c}^1, a_{c}^2, \ldots, t_{c}^n \}\) where \(t_c^i\) represents the \(i^{th}\) sentence in the transcript. Therefore, each sentence will be converted into a vector representation:
\begin{equation}
e_{tc}^i = \text{SimCSE}(t_c^i).
\end{equation}

We obtain the corresponding text embeddings given by \(E_{tc} = \{e_{tc}^1, e_{tc}^2, \ldots, e_{tc}^n \}\). 
The shape of the \(E_{tc}\) is 520 $\times$ 768, where 520 is the maximum number of sentences amongst all data samples and 768 is the dimension of the output of SimCSE. Earnings conference calls with less than 520 sentences ($n < 520$) have been zero-padded for uniformity in input matrix size. 
 
Same with (2)-(5), the MHSA is applied to \(E_{tc}\) to get \(T_{tc} = \{t_{tc}^1, t_{tc}^2, \ldots, t_{tc}^n \}\) with dimension 520 $\times$ 768. Then, the average pooling layer is applied to \(T_{tc} \):
\begin{equation}
T_{t} =  \text{AveragePooling}(T_{tc})
\end{equation}

where  \(T_{t}\) denotes the resultant extracted textual feature of size 768. 

\subsubsection{Earnings Conference Call Analyzer.} This component utilizes an LLM to summarize the earnings call, extracting essential and relevant information and converting it into a text feature. We show this process in Figure~\ref{ECC_Transcript_Analysis}. 

In order to effectively summarize key information in lengthy ECC records, we employ a hierarchical summarization strategy. First, the entire document is divided into chunks, and then we use LLM to summarize each chunk individually. These individual summaries are then further summarized through LLM, resulting in a comprehensive summarization paragraph of the entire document. This two-layer approach ensures that both detailed and aggregate information is captured. In further, we use the OpenAI `text-embedding-3-small' text embedding model to generate embeddings \(T_s\) with size 1024 for summarized paragraph:
\begin{equation}
T_s = \text{Embedding}(\text{Concatenated}[chunk\ summaries; summary])
\end{equation}

Next, we aim to extract the most critical sentences containing relevant information from the entire transcript. To achieve this, we first convert the processed chunks into vector representations using the embedding model. Then, with input from financial experts, we design a list of questions about the ECC, which are stored in a 'Question Bank' as a query list (see Appendix ~\ref{ssec:question}). For each question in the 'Question Bank', we retrieve the relevant chunks from the vectors individually. Next, we use the context compressor, in combination with the LLM chat, to quiz each extracted chunk and query on its relevance. This process filters out only the sentences useful for answering the questions. Finally, we synthesize the query-selected sentences into a coherent prompt, to which the large language model will generate responses as the final answers to the queries. These generated answers are considered as the extracted important sentences. We then concatenate all the generated sentences and feed them into a text embedding model to generate the text embedding \(T_f\) with size 1024:
\begin{equation}
T_f = \text{Embedding}(\text{Concatenated}[selected\ sentences;])
\end{equation}

\begin{table*}[h]
\centering
\caption{Performance results on our proposed framework RiskLabs from different baseline models.}
\scalebox{1.0}{
\begin{tabular}{lccccc|c|c}
\toprule
\textbf{Model}   & \textbf{$\overline{MSE}$}         & \textbf{$MSE_3$} & \textbf{$MSE_7$} & \textbf{$MSE_{15}$} & \textbf{$MSE_{30}$} & \textbf{$VaR$} & \text{Multi-Task}\\ \midrule
Classical Method & 0.713 &1.710   & 0.526  & 0.330   & 0.284   & / & $\otimes$          \\
LSTM             & 0.746 &1.970   & 0.459  & 0.320   & 0.235   & / & $\otimes$           \\
MT-LSTM-ATT      & 0.739 &1.983   & 0.435  & 0.304   & 0.233   & / & $\otimes$            \\
HAN              & 0.598 &1.426   & 0.461  & 0.308   & 0.198   & / & $\otimes$            \\
MRDM             & 0.577 &1.371   & 0.420  & 0.300   & 0.217   & / & $\otimes$            \\
HTML             & 0.401 &0.845   & 0.349  & 0.251   & \textbf{0.158} & / &$\checkmark$         \\
GPT-3.5-Turbo   & 2.198 & 2.152             & 1.793             & 2.514              & 2.332              & 0.371      &    $\checkmark$   \\ \midrule
\textbf{RiskLabs} & \textbf{0.324} & \textbf{0.585} & \textbf{0.317} & \textbf{0.233} & 0.171  & \textbf{0.049}      & $\checkmark$ \\ 
\bottomrule
\end{tabular}}
\label{tab:overall_performance}
\end{table*}

\subsection{Time-Series Encoder} 
For our time series input, we capture the VIX values from a 30-day period preceding the ECC release date. To distill meaningful features from this VIX series, we employ a Bidirectional Long Short-Term Memory (BiLSTM) network~\cite{siami2019performance}: $T_v = \text{BiLSTM}(\textbf{VIX})$. In this setup, the BiLSTM is configured with 64 hidden states. Consequently, the output $T_v$, representing the extracted time series features, has a dimensionality of 128, reflecting the bidirectional nature of the BiLSTM.

\subsection{Relevant News Encoder}
News significantly influences stock movements and serves as a valuable indicator of market trends. It encompasses a wide range of information, including macroeconomic indicators, industry trends, and company-specific events such as earnings reports, mergers and acquisitions, regulatory changes, and management shifts. For each stock, we compile relevant news into a single text string, which is then processed by an LLM to generate a news feature $T_n$.

\subsection{Multimodal Fusion and Multi-Task Learning}
The additive fusion method can be seen as learning a new joint representation:
\begin{equation}
    E = w_0 + w_1\cdot T_a + w_2\cdot T_t + w_3\cdot T_f + + T_v + T_n + \epsilon
\end{equation}
We concurrently model volatility prediction and VaR prediction using a multi-task framework. The multi-task prediction module is comprised of two separate single-layer feedforward networks, each responsible for predicting volatility (vol) and Value at Risk (var) values individually.

RiskLabs is a multi-task training model that trains and predicts the volatility of different terms and VaR. The individual stock volatility can be expressed as the natural log of the standard deviation of return prices \(r\) in a window of \({\tau}\) days. We calculate the 3, 7, 15, and 30 trading days volatility using the equation:
\begin{equation}
\label{e:vol}
v_{[d-\tau, d]} = \ln\left(\left(\frac{\sum_{i=0}^{\tau}(r_{d-i}-\bar{r})^{2}}{\tau}\right)^{\frac{1}{2}}\right) 
\end{equation}
The notations used across the paper are discussed herein. Let \(s \in S\) denote a stock, \(c \in C\) be an earnings call for stock s. For each stock, there exist multiple earnings calls c that are held periodically. Each call c can be segmented into a set of \(a_{c}^i \in A_{c}\) audio clips, and corresponding \(t_{c}^i \in T_{c}\) text sentences for \(i \in [1,N]\), where \(N\) is the maximum number of audio clips in a call. For each stock, there exists a daily return denoted by \(r_i = \ \frac {p_i - p_{i-1}}{p_{i-1}}\), where \(p_i\) is the adjusted close price at the end of the day and \=r is the average return over a period of \({\tau}\) days. We aim to develop a predictive regression function \(f(c) \rightarrow v_{[d-\tau, d]}\)

Our second task is predicting the 1-day VaR of the target stock based on the multi-source inputs. The definition of VaR is:
\begin{align*}
    v = F^{-1}(p)
\end{align*}
The $F(\cdot)$ is the cumulative loss distribution, $p$ is the percentile we set, and $v$ is the VaR. From the idea of Quantile Regression, we can have: 
\begin{align*}
    L_{\tau}(y, \hat{y}) = \begin{cases} 
  \tau \cdot (y - \hat{y}) & \text{if } y \geq \hat{y} \\
  (1 - \tau) \cdot (\hat{y} - y) & \text{if } y < \hat{y} 
\end{cases}
\end{align*}

We train
RiskLabs by optimizing multitask loss:
\begin{equation}
    \mathcal{L} = \mu(\sum_{i}(\hat{y_i}-y_i)^2)+(1-\mu)\text{max}(q\times (v-\hat{v}), (1-q)(\hat{v}-v)).
\end{equation}

\section{Experiment and Discussion}
\subsection{Experiment Setup}
We compare our approach to volatility prediction to several important baselines as described in the Appendix~\ref{app:baseline}. We use GPT-4 for the Earnings Conference Call Analyzer and the News-Market Reactions Encoder. Throughout this process, we set the temperature parameter to 0. This ensures that the Large Language Models (LLMs) produce the most predictable responses, which aids in maintaining consistency in our experiments. For the overall training of the framework, we developed the code using PyTorch. Each Multi-Head Attention layer in the network comprises 8 individual heads, and the training process utilized batch sizes $b \in \{2, 4, 8\}$. We use a grid search to determine the optimal parameters and select the learning rate $\lambda$ for Adam optimizer among $\{1e-3, 1e-5, 1e-6, 1e-7\}$. The best hyper-parameters were kept consistent across all experiments, with the exception of the trade-off parameter $\mu$ which varied between the two tasks. We list additional details in Appendix~\ref{app:dataset}.

\subsection{Performance Comparison  (RQ1 \& RQ2)}
The performance of different models in predicting financial risks is detailed in Table~\ref{tab:overall_performance}. This comparison includes baseline models, our proposed RiskLabs framework, and a range of periods (3, 7, 15, and 30 days). Besides forecasting accuracy, we also consider factors like the predicted Value at Risk (VaR) and the multi-tasking capabilities of each model. Notably, the RiskLabs framework outperforms others in prediction accuracy, particularly in short-term and medium-term forecasts, evidenced by the lowest Mean Squared Error (MSE) values. This improvement becomes more pronounced when the RiskLabs framework integrates various data sources such as earnings conference calls, time series, and news feeds, outshining the current state-of-the-art HTML solutions. 
Moreover, the RiskLabs framework demonstrates superior performance in VaR prediction, underscoring the effectiveness of our proposed methodology in offering a more nuanced and comprehensive approach to financial risk prediction. This could be invaluable for investors seeking to make more informed decisions. Nevertheless, the framework's performance in 30-day forecasts lags behind the HTML models, indicating potential areas for further enhancement in long-term risk forecasting with LLM-based solutions.

In addition to comparing the effectiveness of various AI techniques in risk prediction, our study also evaluates the differences in Value at Risk (VaR) predictions made by Large Language Models (LLMs), traditional financial methodologies, and neural network frameworks. This comparison sheds light on each approach's strengths and weaknesses in risk quantification, offering a clearer understanding of how modern LLMs are different from financial and neural network models. Table \ref{tab:Comparison ai and finance model} displays the performance of value at risk prediction between the traditional finance model and the AI techniques. 

\begin{table*}[h]
\centering
\caption{Ablation Study: performance results of different modules.}
\scalebox{1.0}{
\begin{tabular}{lccccc|c}
\toprule
\textbf{Module}                            & \textbf{$\overline{MSE}$}         & \textbf{$MSE_3$} & \textbf{$MSE_7$} & \textbf{$MSE_{15}$} & \textbf{$MSE_{30}$} & \textbf{$VaR$} \\ \midrule
Audio + Text                           & 0.373      & 0.645         & 0.362         & 0.280          & 0.204          & 0.131         \\
Audio + Text + Analysis                 & 0.357     & 0.627          & 0.335         & 0.267         & 0.199          & 0.057         \\
Audio + Text + Analysis + VIX  & \textbf{0.324} & \textbf{0.585} & \textbf{0.317} & \textbf{0.233} & \textbf{0.171}  & \textbf{0.049}            \\
\bottomrule
\end{tabular}}
\label{tab:ablation}
\end{table*}

\begin{table}[htbp]
  \centering
  \caption{Comparison of Value at Risk Predictions: AI Techniques vs. Traditional Financial Methods. The predefined VaR value is 0.05. }
    \begin{tabular}{cc}
    \toprule
    Method & Prediction of VaR \\
    \midrule
    Historical Method & 0.016 \\
    Fully Connected Neural Network & 0.044 \\
    LSTM  & 0.056 \\
    \midrule
    \textbf{RiskLabs} & \textbf{0.049} \\
    \bottomrule
    \end{tabular}%
  \label{tab:Comparison ai and finance model}%
\end{table}%

The predefined VaR value is 0.05, meaning that the closer the model's predictions are to 0.05, the better its performance.
The result above shows that the prediction of VaR via applying the historical method is 0.016, which is significantly below the pre-defined percentile(5\%). It indicates that the historical method overestimated the 95\% VaR benchmark. Tracing back to 2016, we know there was a global financial crisis in 2015, and its effect lasted till the beginning of 2016. On January 20, 2016, the price of crude oil fell below \$27 a barrel; the DJIA index took a roller coaster from down 565 points to down 249 intraday. In February, the YTD (yield to Date) return came to -10.5\%. These events, together with the sequelae of the 2015 stock market crisis, define 2016 as a risky year. Comparing 2016, 2017 will be much better. In January 2017, DJIA achieved a new historical height, landing above 20,000. The stock market experienced a boost with a 25\% growth rate for DJIA, 19\% for S\&P 500, and 28\% for Nasdaq. The market had strong confidence, and the VIX index in 2017 came to its historical lowest point. That explains the reason why the historical method may overestimate the 95\% VaR benchmark, due to this method duplicating the extreme scenarios from 2016 to 2017, which leads to the extra estimation of financial risks. We show more results of the analysis and visualization in Appendix~\ref{app:var}.

Turning our focus to Table~\ref{tab:overall_performance}, we observe that the direct application of LLMs for financial risk prediction is markedly ineffective, akin to making random guesses. This underscores a crucial caution; if LLMs are not utilized appropriately, they might elevate investment risks. Consequently, in response to Research Questions 1 and 2 (RQ1 and RQ2), we conclude:
\begin{itemize}
\item Utilizing LLMs through simple prompt instructions for direct financial risk prediction is ineffectual and potentially hazardous, increasing investment risks.
\item \textbf{LLM is a bad trader/predictor, but it's a helpful assistant.} While LLMs alone may not be reliable for direct risk prediction, they can serve as valuable tools in collating and analyzing diverse financial data. This processed information, when fed into sophisticated deep learning models, significantly enhances AI's capability in forecasting financial risks, thus positioning LLMs as beneficial assistants rather than standalone predictors.
\end{itemize}

\subsection{Comparison Across Modules (RQ3)}

In addition to overall performance, we evaluate the
impact of different data combinations on predictive performance to better understand the relative
contributions of data sources and each sub-module. We designed the ablation study as follows:
\begin{itemize}
    \item \textbf{Audio + Text:} We design a comparison experiment starting with only using earnings conference calls. These calls are crucial in financial analysis as they are where executives of publicly traded companies discuss the company's financial results for a specific period. They offer valuable insights into the company's performance, strategic initiatives, and future projections. Investors highly regard these calls for the direct access they provide to the company's leadership, offering in-depth financial data, operational updates, and forward-looking statements essential for informed investment decisions. In our approach, we initially concentrate on processing both the audio and textual content from these earnings conference calls using Multi-Head Self-Attention to extract salient features. Subsequently, we integrate these extracted features to directly predict financial outcomes.
     \item \textbf{Audio + Text + Analysis:} Then, we integrate the earnings conference calls analysis results into the prediction process. This is also the complete earnings conference call encoder in RiskLabs. This experiment will help us to verify whether the earnings conference calls analysis results are useful in helping to improve the model prediction performance.
     \item \textbf{Audio + Text + Analysis + VIX:} Next, we incorporate time series information into the model as well. We will determine whether the introduction of time series information helps the model to make predictions by observing changes in predictive performance.
\end{itemize}

 As shown in Table~\ref{tab:ablation}, the Audio + Text' combination outperforms the baseline HTML model for 3-day forecasts. For longer periods—7, 15, and 30 days—RiskLabs achieves predictions closely aligned with HTML while still surpassing other baselines. Notably, with just Audio + Text', our model is more streamlined than HTML, benefiting from a pre-trained model with extensive parameters that effectively map text and audio to vector representations. These vectors are processed through a multi-head self-attention mechanism for feature extraction, demonstrating the value of large-scale models in enhancing AI-driven risk prediction. Further improvements were observed upon integrating earnings conference call analysis and time-series data, leading to incremental gains in RiskLabs' performance, especially for medium- to long-term forecasts (7, 15, and 30 days). These results suggest that earnings conference calls, as essential references for investors, have a more substantial impact on short-term risk volatility, while diverse information sources are crucial for improving accuracy in long-term forecasts. These findings underscore the contributions of each component within the RiskLabs framework, culminating in a robust predictive model.

\subsection{Challenge and Opportunity}
Our experimental findings demonstrate that leveraging large language models to integrate diverse
information sources can significantly enhance the capability of our framework in predicting financial
risks. With this in mind, we’ve expanded our data sources to include variables that could influence
market volatility, such as news disseminated through social media. As detailed in Section 2.4, we
utilize LLM to collect and analyze daily financial news. Initial small-scale experiments indicate
that this addition can further improve the model’s performance. However, scaling up has introduced
challenges: 1) the variable quality of news sources, with some containing misinformation, potentially
introduces noise into our news database and affects audio model predictions; 2) there is a need to test
the module’s effectiveness across a broader data spectrum.
To address these challenges, we are implementing more nuanced steps to filter out low-quality
news. Additionally, we are amassing newer data samples to enlarge our dataset, allowing for more
comprehensive validation of our model’s efficacy. Furthermore, we will implement: a) Bayes-VaR method provides a probabilistic framework that allows for the incorporation of prior knowledge and the updating of
beliefs as new evidence is presented. b) News-Market-Reactions Encoder c) "Dynamic Moving
Time Window" and d) a "Time Decay Hyper-parameter" in RiskLabs. These features
will enable more flexible training and forecasting on a daily basis, thus offering investors timelier and
more accurate risk assessments. We described ongoing tasks as follows.

\subsubsection{\textbf{Extract Relationship Among Multiple Response Variable Using Vector Auto-regression-based Method}}

The task of RiskLabs is to predict multiple risk metrics in the meantime. Relationships will exist among them.
Therefore, by identifying and modeling the relationships between them, we can significantly improve the accuracy of our predictions. 
This method shifts the focus from treating each metric as an isolated entity to understanding them as part of a complex system, where the dynamics between variables can provide critical insights for more precise risk assessment. 

To achieve this, we attempt to find the relationship that links the two different stages: the $m$-day before and the $m$-day after.  We will use the VAR(Vector Auto-Regression Model) to capture the linear relationship between the volatility of different terms. We set four different "$m$-day volatility"(3-day volatility, 7-day volatility, 15-day volatility, and 30-day volatility). We define the $\sigma_{-m,t}$ as the "$m$-day future volatility" at time $t-m$; the calculation method traces back $m-day$ from today and calculates the $m-day$'s standard deviation of returns. 

Volatility typically shows the cluster characteristics, which means that the change transmission pattern to the next stage tends to be similar to the last transmission pattern. These characteristics indicate that we could use the results from the close historical data as indicators or predictors to predict the next stage. We assume that the correlation between the volatility among different time scales will affect each other. We also measure the correlation by estimating the coefficient matrix and the information included in the historical data.

To estimate the coefficient of the model, we use the Bayesian Methodology by estimating the posterior distribution of the coefficient. If we have a general linear regression $Y = X\theta$, here $\theta$ is the coefficient that we need to estimate. We can have the:
\begin{equation}
    P(\theta|Y,X) = \frac{P(Y|\theta, X)P(\theta)}{P(Y|X)} = \frac{P(Y|\theta, X)P(\theta|X)}{\int_{-\infty}^{\infty}P(Y|X, \theta)p(\theta)d\theta}
    \label{Bayesian 2}
\end{equation}
Here, $P(\theta)$(or $P(\theta|X)$) is the prior distribution that we could guess from the existing research or common knowledge. We also suppose that the independent variable is independent with the regression coefficient; $P(Y|\theta, X)$ is the Likelihood Estimation from the data, and $P(Y|X)$ is the marginal distribution. Usually, we could view $P(Y|X) = \int_{-\infty}^{\infty}P(Y|X, \theta)p(\theta)d\theta$ as a constant because we could have the real value from the true dataset. From the Equation \ref{Bayesian 2}, we could have the following relation:
\begin{equation}
    P(\theta|Y,X) \propto P(Y|\theta, X)P(\theta|X)
\end{equation}
If we can determine the prior and the likelihood function, we can estimate the posterior distribution. We assume that the prior distribution $P(\theta)$ satisfies the Normal distribution based on the simple linear regression. Subsequently, we consider the Likelihood $P(Y|\theta, X)$. In the empirical study from \cite{andersen2001distribution}, the unconditional distribution of the realized historical distribution of the volatility is highly right-skewed. Besides, \cite{cizeau1997volatility} tested the distribution historical volatility of S\&P 500 stock from 1984 to 1996 and demonstrated that the pattern follows the log-normal distribution. Therefore, we took the log form of the original data. Given the preparation, we build the VAR model as follows:
\begin{align*}
    \left\{
    \begin{array}{l}
        log(\sigma_{3,t}) = \alpha_{3} + \beta_{1,1}log(\sigma_{-3,t}) + \beta_{1,2}log(\sigma_{-7,t}) + \beta_{1,3}log(\sigma_{-15,t}) \\
        + \beta_{1,4}log(\sigma_{-30,t}) + u_{3,t}\\
        log(\sigma_{7,t}) = \alpha_{7} + \beta_{2,1}log(\sigma_{-3,t}) + \beta_{2,2}log(\sigma_{-7,t})\\ 
        + \beta_{2,3}log(\sigma_{-15,t}) + \beta_{2,4}log(\sigma_{-30,t}) + u_{7,t} \\
        log(\sigma_{15,t}) = \alpha_{15} + \beta_{3,1}log(\sigma_{-3,t}) + \beta_{3,2}log(\sigma_{-7,t}) \\
        + \beta_{3,3}log(\sigma_{-15,t}) + \beta_{3,4}log(\sigma_{-30,t}) + u_{15,t} \\
        log(\sigma_{30,t}) = \alpha_{30} + \beta_{4,1}log(\sigma_{-3,t}) + \beta_{4,2}log(\sigma_{-7,t}) \\
        + \beta_{4,3}log(\sigma_{-15,t}) + \beta_{4,4}log(\sigma_{-30,t}) + u_{30,t}
    \end{array}
    \right.
\end{align*}

Here, $\sigma_{m,t}$ stands for the "$m$-day" future volatility at time $t$; $u_{m,t}$ is a White noise term, which $u_{m,t} \sim N(0,1)$; $\beta_{i,j}$ represents the linear relationship, which is the coefficient matrix of the VAR model, $\alpha_m$, which is the intercept term of the VAR model.

We apply the Monte Carlo Markov Chain(MCMC) algorithm to obtain the posterior distribution of the coefficient. MCMC is a method that is based on Bayesian Estimation. It is mainly used to predict the posterior distribution in a probability space by sampling. The Markov Chain follows the equation:
\begin{align*}
    P(X_{t+1}|X_1,X_2,...,X_t) = P(X_{t+1}|X_t)
\end{align*}
This could be explained by the fact that the probability of transforming only depends on the former statement. And based on the convergence theorem of Markov Chain, suppose the Markov Chain converges to one probability at $n$th step: $\pi_n(x)\rightarrow \pi(x)$, and $X_n\sim \pi(x)$, $X_{n+1}\sim \pi(x)$, $...$, we could say that $X_n$, $X_{n+1}$, $...$ are i.i.d random variables.

We attempt to calculate the expectation of the posterior distribution by sampling. However, the posterior distribution is challenging to measure. To achieve this, we introduce the MCMC method to create a Markov Chain to achieve a stationary distribution and make it our desired posterior distribution. For MCMC, Metropolis-Hastings introduced a system to find a Markov Chain. For the model with many parameters to estimate, another alternative way is to use the Gibbs Sampling to update the estimated parameters one by one.

Regarding testing the MCMC process, we use several matrices, such as Monte Carlo standard errors (MCSE). There are two types, which measure the standard error of the mean and the standard error of the standard deviation of the chains. Also, because we did not generate the independent samples, the simulated samples are correlated. We want to know how many theoretically independent samples we drew and create the effective sample size (ESS), which measures the number of effectively independent samples we draw. Besides, we need to test if we achieve the stationary of our chains. Gelman-Rubin R-hat statistic($\hat{R}$) gave us a measure. We measure $\hat{R}$ by:
\begin{align*}
    \hat{R}=\frac{Variance\ between\ Chains}{Variance\ within\ Chains}
\end{align*}
Usually, $\hat{R}$ should close to 1 and less than 1.01 otherwise it doesn't achieve the stationary with the Markov Chain.

\begin{table}[htbp]
  \centering
  \caption{Bayesian-VAR results by applying the MCMC method}
  \resizebox{\linewidth}{!}{
    \begin{tabular}{cccccccccccc}
    \toprule
    Dependent & Independent & N     & Mean  & Sd    & Hdi\_3\% & Hdi\_97\% & mcse\_mean & mcse\_sd & ess\_bulk & ess\_tail & r\_hat \\
    \midrule
    \multirow{5}[2]{*}{3-Day} & Intercept & 250   & -2.571 & 0.364 & -3.275 & -1.908 & 0.004 & 0.003 & 8342.0 & 7461.0 & 1.0 \\
          & 3-Day & 250   & 0.070 & 0.058 & -0.044 & 0.174 & 0.001 & 0.000 & 9345.0 & 8076.0 & 1.0 \\
          & 7-Day & 250   & 0.118 & 0.116 & -0.107 & 0.329 & 0.001 & 0.001 & 8512.0 & 7501.0 & 1.0 \\
          & 15-Day & 250   & 0.017 & 0.146 & -0.249 & 0.297 & 0.002 & 0.001 & 7259.0 & 7407.0 & 1.0 \\
          & 30-Day & 250   & 0.203 & 0.143 & -0.079 & 0.465 & 0.002 & 0.001 & 7072.0 & 7532.0 & 1.0 \\
    \midrule
    \multirow{5}[2]{*}{7-Day} & Intercept & 250   & -2.748 & 0.251 & -3.207 & -2.265 & 0.003 & 0.002 & 8684.0 & 7809.0 & 1.0 \\
          & 3-Day & 250   & 0.042 & 0.040 & -0.033 & 0.118 & 0.000 & 0.000 & 8600.0 & 7890.0 & 1.0 \\
          & 7-Day & 250   & 0.181 & 0.079 & 0.033 & 0.330 & 0.001 & 0.001 & 7281.0 & 7845.0 & 1.0 \\
          & 15-Day & 250   & -0.078 & 0.100 & -0.270 & 0.104 & 0.001 & 0.001 & 7428.0 & 7426.0 & 1.0 \\
          & 30-Day & 250   & 0.142 & 0.098 & -0.035 & 0.333 & 0.001 & 0.001 & 7308.0 & 7386.0 & 1.0 \\
    \midrule
    \multirow{5}[2]{*}{15-Day} & Intercept & 250   & -3.350 & 0.207 & -3.729 & -2.960 & 0.002 & 0.002 & 9410.0 & 8082.0 & 1.0 \\
          & 3-Day & 250   & 0.039 & 0.034 & -0.025 & 0.103 & 0.000 & 0.000 & 10471.0 & 8203.0 & 1.0 \\
          & 7-Day & 250   & 0.068 & 0.067 & -0.058 & 0.194 & 0.001 & 0.001 & 8018.0 & 7923.0 & 1.0 \\
          & 15-Day & 250   & -0.026 & 0.085 & -0.182 & 0.134 & 0.001 & 0.001 & 7053.0 & 7467.0 & 1.0 \\
          & 30-Day & 250   & 0.006 & 0.083 & -0.149 & 0.165 & 0.001 & 0.001 & 7373.0 & 8124.0 & 1.0 \\
    \midrule
    \multirow{5}[2]{*}{30-Day} & Intercept & 250   & -3.604 & 0.164 & -3.904 & -3.280 & 0.002 & 0.001 & 9388.0 & 8508.0 & 1.0 \\
          & 3-Day & 250   & 0.015 & 0.027 & -0.034 & 0.067 & 0.000 & 0.000 & 9676.0 & 8515.0 & 1.0 \\
          & 7-Day & 250   & 0.108 & 0.052 & 0.006 & 0.200 & 0.001 & 0.000 & 7919.0 & 7510.0 & 1.0 \\
          & 15-Day & 250   & 0.005 & 0.065 & -0.115 & 0.129 & 0.001 & 0..001 & 7004.0 & 7650.0 & 1.0 \\
          & 30-Day & 250   & -0.140 & 0.065 & -0.263 & -0.020 & 0.001 & 0.001 & 7145.0 & 7494.0 & 1.0 \\
    \bottomrule
    \end{tabular}}%
  \label{tab:regression results}%
\end{table}%

In order to validate this methodology, we choose one of the stocks "TWTR," in our database as a case study, and then output the Bayesian regression results. Here, we used 250 days of data(from 2016-02-22 to 2017-02-15) as the model training period. 
Table \ref{tab:regression results} shows the training result. We observe that the $\hat{R}$ of all parameters is equal to 1, and the $ess\_tail$ value and $ess\_bulk$ are all over 7000. Typically, there is no specific argument about the number of ESS. \cite{kruschke2014doing} in the book mentioned that we need more ESS samples for the interested parameters. In the meantime, we also need to balance the number of ESS data and model training costs.~\cite{vehtari2021rank} in the paper recommends having the ESS number over 400, which should be dependent on practical experience and model, to get a stable Monte Carlo standard error.

\begin{table}[htbp]
  \centering
  \caption{Statistics Results about AEP(Absolute Error Percentage) of the Experiment}
  \resizebox{\linewidth}{!}{
    \begin{tabular}{ccccccccccc}
    \toprule
    Variable & N     & Mean  & SD    & Skewness & Kurtosis & 5\% Percentile & 25\% Percentile & 50\% Percentile & 75\% Percentile & 95\% Percentile \\
    \midrule
    3-Day & 100   & -0.13167 & 0.08377 & -0.42741 & -0.70133 & -0.26854 & -0.19993 & -0.13011 & -0.07004 & -0.01447 \\
    7-Day & 100   & -0.08501 & 0.05202 & -0.37466 & -0.43682 & -0.17094 & -0.11859 & -0.08499 & -0.04545 & -0.00960 \\
    15-Day & 100   & -0.07950 & 0.05310 & -0.23159 & -1.37432 & -0.16238 & -0.11840 & -0.07607 & -0.03119 & -0.00952 \\
    30-Day & 100   & -0.04195 & 0.03775 & -1.75213 & 3.12077 & -0.11573 & -0.05777 & -0.03011 & -0.01555 & -0.00600 \\
    \bottomrule
    \end{tabular}}%
  \label{tab:AEP results}%
\end{table}%

We used the rolling window(see Figure \ref{Fig: Rolling window}) to make a rough simulation only based on the historical data of different volatility terms. We chose 250 days(Approximately one year) as the training set window for the next-day prediction and iterated 100 times. The result is shown in Table \ref{tab:AEP results}. This table measures the distribution of AEP(Absolute Error Percentage), which is calculated by the following formula:
\begin{align*}
    AEP = \frac{|\hat{y}-y|}{y}
\end{align*}
Here, the $\hat{y}$ is the estimated value from the Bayesian VAR model, and $y$ is the true value. From Table \ref{tab:AEP results}, we can see that the average estimation bias will become smaller when we estimate the longer volatility term, the same as the standard deviation of AEP. The negative skewness tells us that the peak of the AEP distribution is close to the y-axis(Because the mean is negative, as we calculated). In general, this experiment shows that using the VAR model based on historical data to predict the future term should be possible and could maintain some accuracy.

\subsubsection{\textbf{News-Market-Reactions Encoder}}
News plays an important role in influencing stock movements, serving as a powerful indicator of market trends. It contains a wealth of information, ranging from macroeconomic indicators and industry trends to company-specific news such as earnings reports, mergers and acquisitions, regulatory changes, and management shifts. 

News carries several key features that commonly impact its reception and influence on its audience.
Firstly, news is most impactful when it is fresh. Information that is current and up-to-date is more likely to influence decisions and perceptions than outdated news. 
Secondly, similar news tends to produce comparable impacts on market reactions. When news items share key characteristics—such as subject matter, sentiment, and relevance to investors—they often trigger similar responses among market participants.  
For instance, announcements of unexpected earnings surpassing market forecasts typically lead to positive stock price movements. Conversely, news of regulatory setbacks or legal challenges can prompt a downturn in stock values. This patterned response largely stems from investors collectively interpreting news through the lens of their past experiences and established market precedents, leading to the development of conditioned reactions to specific types of news. 
Consequently, understanding the similarities in news items can offer insights into potential market reactions. 
However, given the volume of news a company may receive in a day, and considering that each piece of news partially influences the future stock price, it's essential to assess the entire context. To account for the cumulative effect and interrelations among different news items, we analyze them collectively, treating all news related to a company within a single day as a unified analytical unit. 

\begin{figure*}[h]
    \centering
    \includegraphics[width=0.8\textwidth]{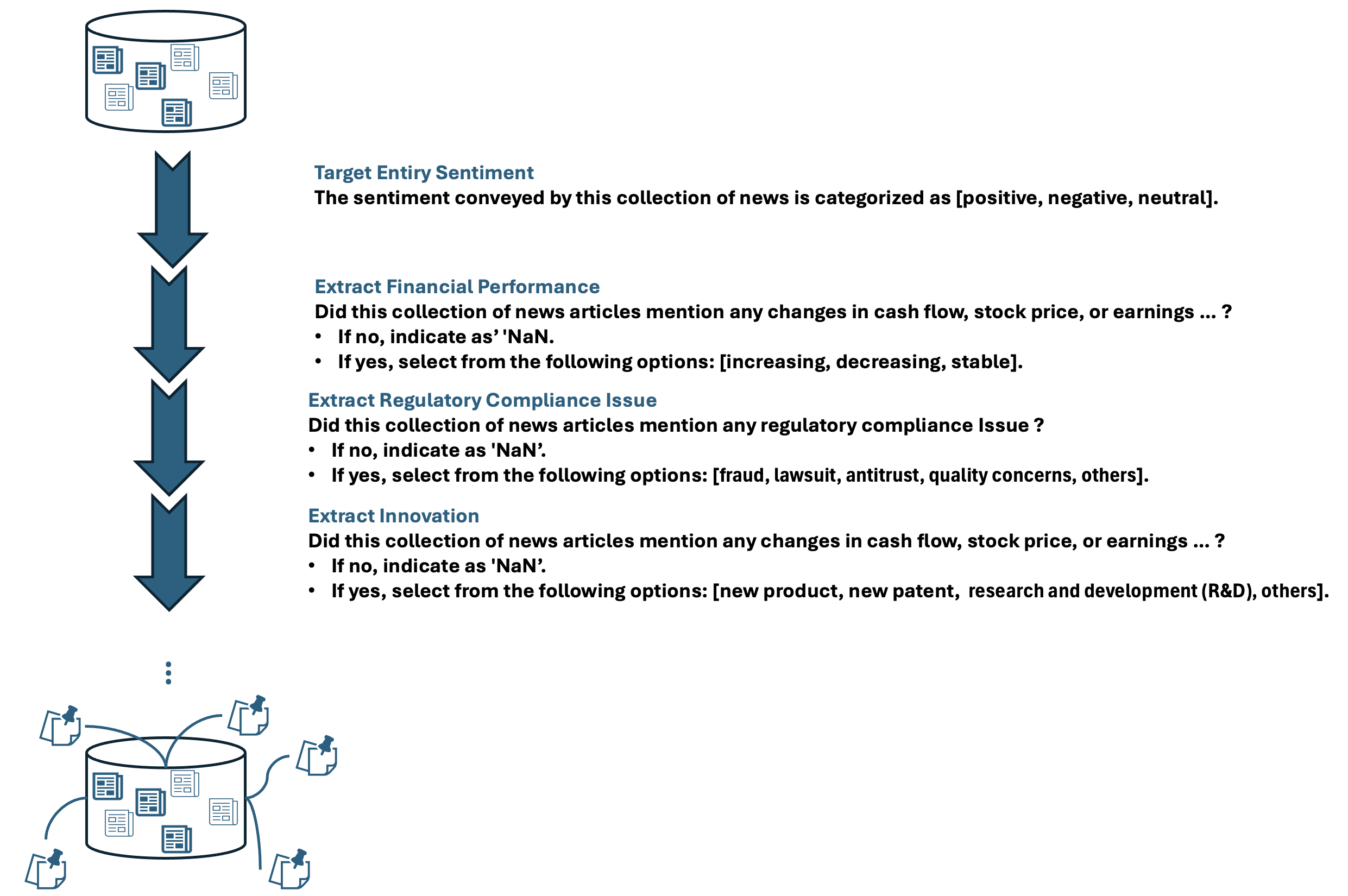}
    \caption{This figure illustrates the pipeline for enriching the news information. First, the pipeline will analyze the sentiments from the target news. Then, based on the binary questions bank designed for different topics, Pipeline can extract the information and answer these questions. Finally, if the pipeline could capture the signal of a specific topic,  it would also give feedback on the potential market response.}
    \label{Enrich_News_Pipeline}
\end{figure*}

Initially, we collect news pertaining to a particular company from the three days prior to our target trading day, including the market responses for each day. This compilation allows a Large Language Model (LLM) to examine the data and deduce how the news might influence the stock's performance in the subsequent days. Subsequently, our strategy involves identifying historical dates on which the news profiles closely resemble those on the day before the intended trading day. This comparative analysis aims to understand the potential market reactions based on similar past events, thereby informing our trading strategies with a nuanced perspective on news impact.

Based on these news features, we propose a news-market reactions encoder composed of two main components.  Firstly, we collect news for a specific company from the three days prior to our target trading day, along with the market reactions for each day. We then use an LLM to analyze the data and identify how the news might affect the stock's performance in the days that follow.
Secondly, we attempt to identify the historical date whose news is similar to that of the intended trading date. In this scenario, we face the challenge of assessing similarity once more. Given that we are comparing two groups of news, the likelihood of identical combinations of news reoccurring is extremely small, which may lead to less accurate retrieval results.

To address this problem, we designed an enriched news pipeline to obtain attributes associated to the news and then attach this news to news group. Specifically, the enriched news pipeline comprises several key steps to analyze news content effectively. 

After processing all news in our database through this pipeline, we assign these identified attributes to each news group as metadata. This metadata becomes instrumental in locating news similar to that of the target trading days. Rather than directly comparing two groups of news, we initially compare the metadata between them. This process helps us identify the top k news groups in the historical record that share similar metadata attributes. Subsequently, we assess the similarity between these top k groups and the news groups from the target trading date. 

\begin{figure*}[h]
    \centering
    \includegraphics[width=0.4\textwidth]{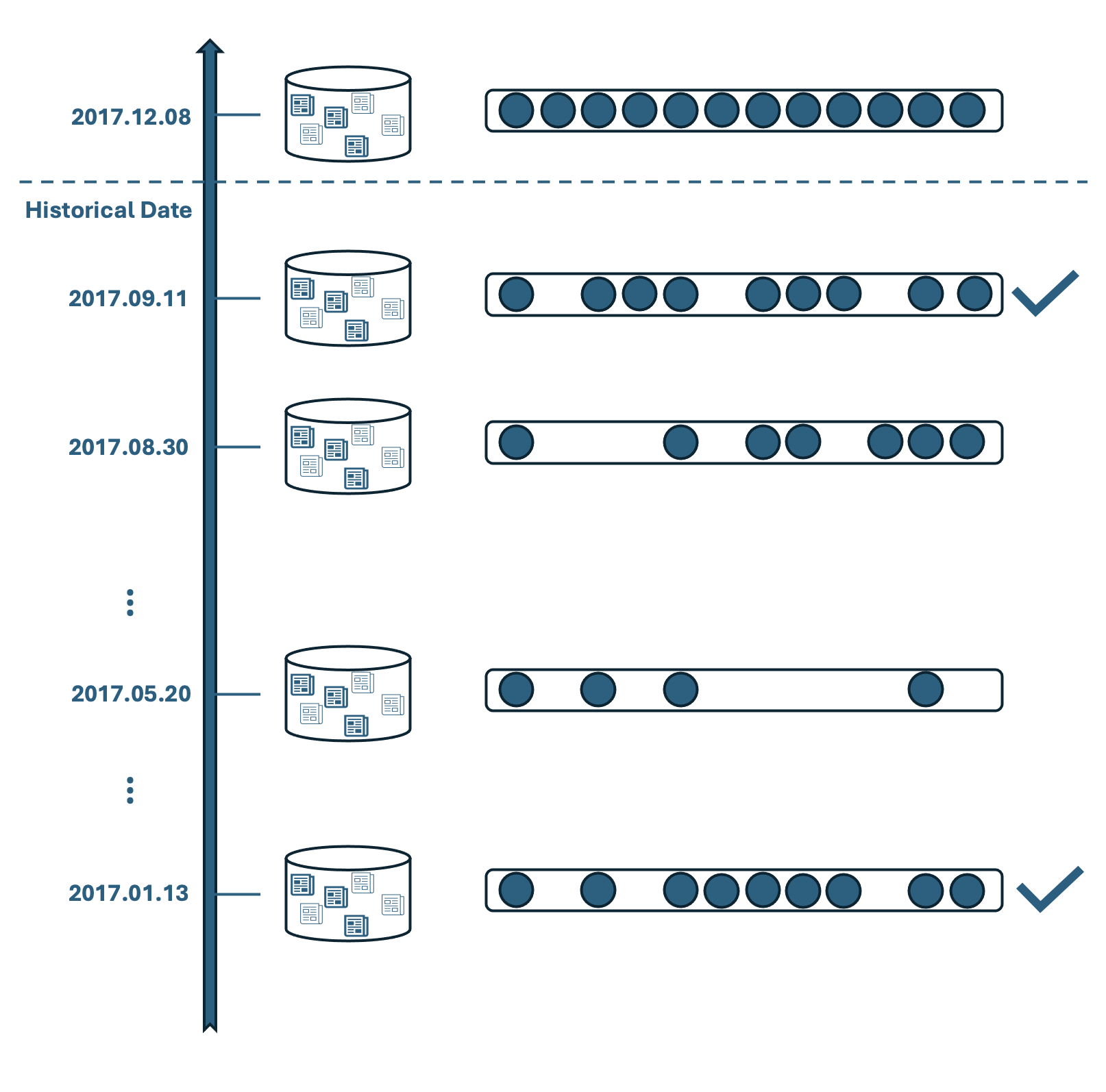}
    \caption{This diagram illustrates the process by which the News Analyzer assesses similarities across various news collections. As news items pass through the enrichment pipeline, they are tagged with multiple attributes. Each circle in the figure represents one of these attributes. To identify similar news collections from historical data, the analyzer starts by comparing these attributes and filtering out certain ones. Subsequently, it evaluates the similarity among the remaining attributes to determine the connections between different news collections.}
    \label{Fig:FindSimilarNews}
\end{figure*}

This method significantly streamlines the process of finding similar news groups, enabling us to efficiently draw parallels. With these similarities established, we can make informed predictions about the potential impact of these news groups on the stock market in the days following the target trading date, based on the market reactions to historically similar news.

In this way, it is efficient to identify the two groups that are similar, then we can make an inference for the target trading date about how these results could impact the following stock market based on the historically similar news and its corresponding market reactions.

\subsubsection{\textbf{Time Decay Hyper-parameter and Dynamic Moving Time Window Training}}
In previous sections, we examined the model inputs: (1) earnings conference call transcripts, (2) historical time series data, and (3) news articles. However, the availability frequency of these inputs varies. News and historical time series data are available on a daily basis, whereas earnings conference calls are released on specific dates when a company presents its earnings report. 

\begin{figure*}[h]
    \centering
    \includegraphics[width=0.5\textwidth]{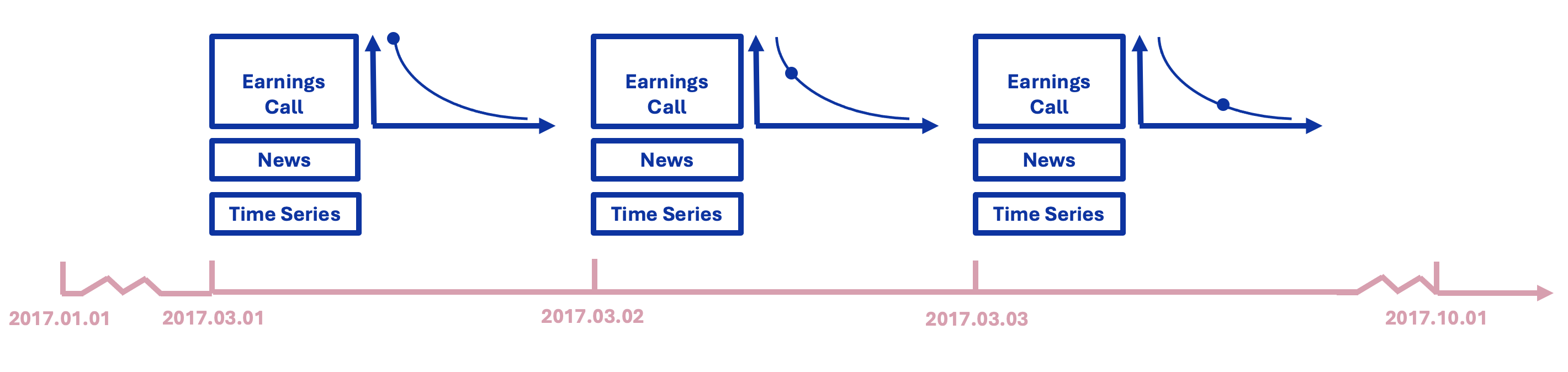}
    \caption{This figure illustrates the methodology for modeling the impact of earnings conference calls on the stock market when earnings conference calls are not available. It uses a curve to represent the time decay effect of an earnings conference call's influence over time. When an earnings conference call is initially issued, its impact on the market is at its peak. Over time, this influence gradually diminishes.}
    \label{Fig: Time Decaying}
\end{figure*}

Consequently, the input from earnings conference calls may be absent on some training days. Ignoring this input entirely would be imprecise, as the information from an earnings conference call can continue to affect stock price movements after its publication. To incorporate the continuous influence of earnings conference call information even on days without new input, we will introduce a hyper-parameter to measure the decaying speed of its relevance. We will apply the Exponential Decay function:
\begin{align*}
    I(t) = I(0)e^{-\lambda t}
\end{align*}
In the equation, $\lambda$ is the hyper-parameter that measures the decay rate. $t$ is the time to the last earnings conference call release date. When the $\lambda$ is fixed, the impact of the earnings conference call will still decrease as $t$ gets larger, which satisfies our common sense. Figure \ref{Fig: Time Decaying} shows the daily data inputs, for days lacking a new earnings conference call, we still incorporate the information from the most recent call, adjusted by $\lambda$ to reflect its diminishing impact speed over time. 

Another consideration related to timing is the impact of input data on the market across time.
RiskLabs takes time series data and news as inputs. The relevance of these inputs to the response variables is strongly dependent on the timing of the data. If we solely train a model without further updates and then use it to make predictions for dates far away the training period, the model's predictive performance may decline. This decrease in accuracy arises because the model becomes less relevant of later dates.

\begin{figure*}[h]
    \centering
    \includegraphics[width=0.5\textwidth]{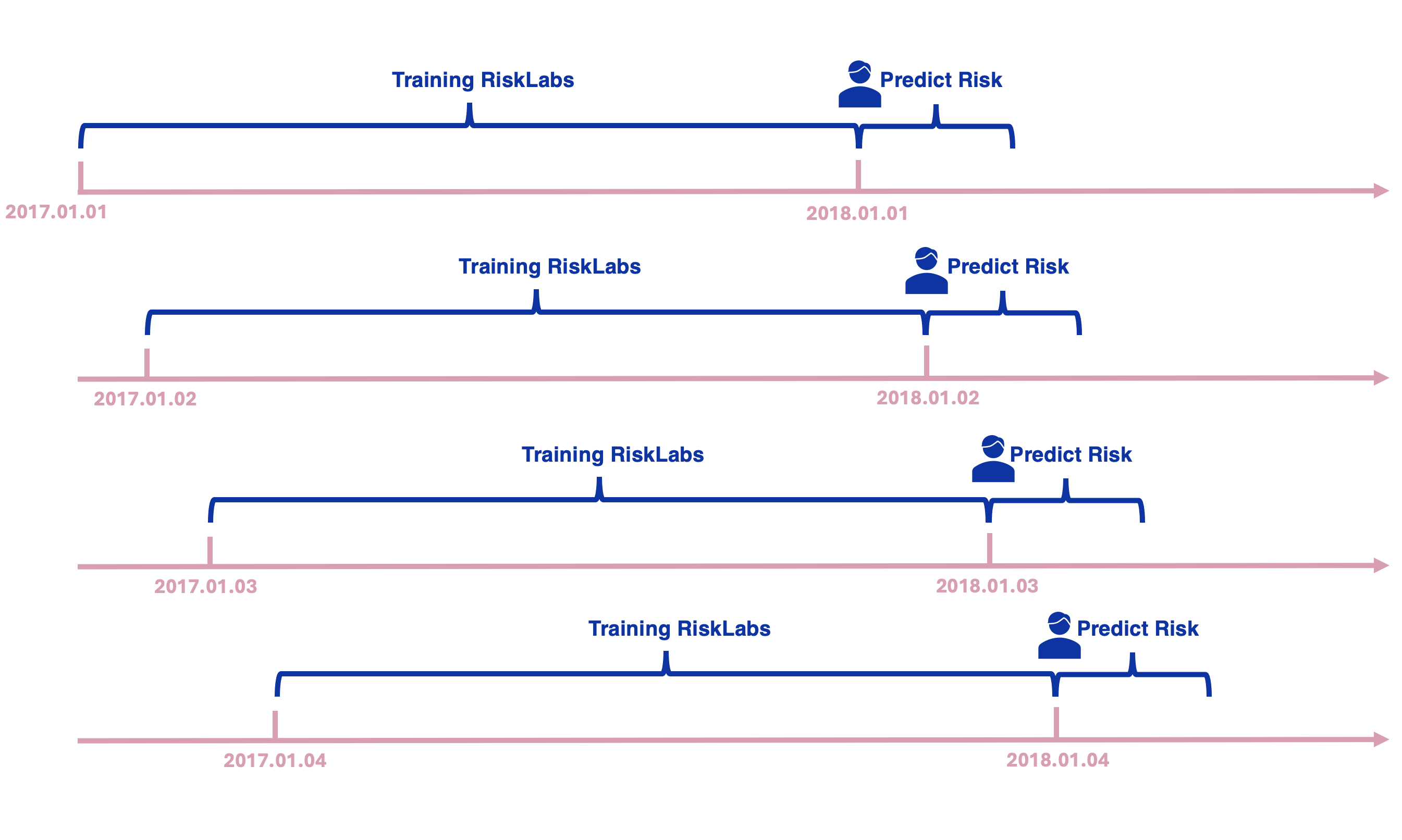}
    \caption{This figure provides a visual representation of the Rolling Window methodology in action. By establishing a fixed window length, the approach systematically progresses the training set forward, day by day, in alignment with the passage of time.}
    \label{Fig: Rolling window}
\end{figure*}

Hence, updating the model at a high frequency can continuously adjust the model's parameters to maintain its sensitivity to the latest market trends and fluctuations. To achieve this, we use a dynamic moving time window method for training. Specifically, we use a fixed window of historical data up to the target trading day to train our model, which is then used only to predict the response variable for that particular day. Once the day has passed, we shift the window to include the newly added day and proceed to train a new model for the next target trading day(See Figure \ref{Fig: Rolling window}). This iterative process ensures our model remains adaptive and up-to-date with the latest market data.

\section{Conclusion}
In this study, we explored the utilization of LLMs in predicting financial risks and introduced the RiskLabs framework. This innovative framework employs LLMs to systematically organize and analyze diverse financial data types and sources, augmenting deep learning models in financial risk prediction. Central to RiskLabs are specialized modules: the Earnings Conference Call Encoder, the Time-Series Encoder, and the News-Market Reactions Encoder, each designed to process specific financial data. These encoders collectively facilitate the merging of various data features for robust multi-task financial risk forecasting. The empirical findings from our study led to several key insights: 1) The RiskLabs framework demonstrates a high efficacy in predicting financial risks, confirming its potential as a valuable tool in this domain. 2) While LLMs in isolation may not yield effective financial risk predictions, their strategic application in processing relevant financial data significantly enhances the predictive power of deep learning models. 3) Ablation studies further affirm that each individual module of RiskLabs meaningfully contributes to the accuracy of the final risk predictions. Overall, our research underscores the transformative potential of LLMs in financial risk assessment, marking a significant step forward in the application of AI in finance.


\bibliographystyle{ACM-Reference-Format}
\bibliography{sample-base}


\begin{thebibliography}{33}


\ifx \showCODEN    \undefined \def \showCODEN     #1{\unskip}     \fi
\ifx \showDOI      \undefined \def \showDOI       #1{#1}\fi
\ifx \showISBNx    \undefined \def \showISBNx     #1{\unskip}     \fi
\ifx \showISBNxiii \undefined \def \showISBNxiii  #1{\unskip}     \fi
\ifx \showISSN     \undefined \def \showISSN      #1{\unskip}     \fi
\ifx \showLCCN     \undefined \def \showLCCN      #1{\unskip}     \fi
\ifx \shownote     \undefined \def \shownote      #1{#1}          \fi
\ifx \showarticletitle \undefined \def \showarticletitle #1{#1}   \fi
\ifx \showURL      \undefined \def \showURL       {\relax}        \fi
\providecommand\bibfield[2]{#2}
\providecommand\bibinfo[2]{#2}
\providecommand\natexlab[1]{#1}
\providecommand\showeprint[2][]{arXiv:#2}

\bibitem[Ahbali et~al\mbox{.}(2022)]%
        {ahbali2022identifying}
\bibfield{author}{\bibinfo{person}{Noujoud Ahbali}, \bibinfo{person}{Xinyuan Liu}, \bibinfo{person}{Albert Nanda}, \bibinfo{person}{Jamie Stark}, \bibinfo{person}{Ashit Talukder}, {and} \bibinfo{person}{Rupinder~Paul Khandpur}.} \bibinfo{year}{2022}\natexlab{}.
\newblock \showarticletitle{Identifying corporate credit risk sentiments from financial news}. In \bibinfo{booktitle}{\emph{Proceedings of the 2022 Conference of the North American Chapter of the Association for Computational Linguistics: Human Language Technologies: Industry Track}}. \bibinfo{pages}{362--370}.
\newblock


\bibitem[Andersen et~al\mbox{.}(2001)]%
        {andersen2001distribution}
\bibfield{author}{\bibinfo{person}{Torben~G Andersen}, \bibinfo{person}{Tim Bollerslev}, \bibinfo{person}{Francis~X Diebold}, {and} \bibinfo{person}{Heiko Ebens}.} \bibinfo{year}{2001}\natexlab{}.
\newblock \showarticletitle{The distribution of realized stock return volatility}.
\newblock \bibinfo{journal}{\emph{Journal of financial economics}} \bibinfo{volume}{61}, \bibinfo{number}{1} (\bibinfo{year}{2001}), \bibinfo{pages}{43--76}.
\newblock


\bibitem[Baevski et~al\mbox{.}(2020)]%
        {baevski2020wav2vec}
\bibfield{author}{\bibinfo{person}{Alexei Baevski}, \bibinfo{person}{Yuhao Zhou}, \bibinfo{person}{Abdelrahman Mohamed}, {and} \bibinfo{person}{Michael Auli}.} \bibinfo{year}{2020}\natexlab{}.
\newblock \showarticletitle{wav2vec 2.0: A framework for self-supervised learning of speech representations}.
\newblock \bibinfo{journal}{\emph{Advances in neural information processing systems}}  \bibinfo{volume}{33} (\bibinfo{year}{2020}), \bibinfo{pages}{12449--12460}.
\newblock


\bibitem[Chung et~al\mbox{.}(2014)]%
        {chung2014empirical}
\bibfield{author}{\bibinfo{person}{Junyoung Chung}, \bibinfo{person}{Caglar Gulcehre}, \bibinfo{person}{KyungHyun Cho}, {and} \bibinfo{person}{Yoshua Bengio}.} \bibinfo{year}{2014}\natexlab{}.
\newblock \showarticletitle{Empirical evaluation of gated recurrent neural networks on sequence modeling}.
\newblock \bibinfo{journal}{\emph{arXiv preprint arXiv:1412.3555}} (\bibinfo{year}{2014}).
\newblock


\bibitem[Cizeau et~al\mbox{.}(1997)]%
        {cizeau1997volatility}
\bibfield{author}{\bibinfo{person}{Pierre Cizeau}, \bibinfo{person}{Yanhui Liu}, \bibinfo{person}{Martin Meyer}, \bibinfo{person}{C-K Peng}, {and} \bibinfo{person}{H~Eugene Stanley}.} \bibinfo{year}{1997}\natexlab{}.
\newblock \showarticletitle{Volatility distribution in the S\&P500 stock index}.
\newblock \bibinfo{journal}{\emph{Physica A: Statistical Mechanics and its Applications}} \bibinfo{volume}{245}, \bibinfo{number}{3-4} (\bibinfo{year}{1997}), \bibinfo{pages}{441--445}.
\newblock


\bibitem[Franses and Van~Dijk(1996)]%
        {franses1996forecasting}
\bibfield{author}{\bibinfo{person}{Philip~Hans Franses} {and} \bibinfo{person}{Dick Van~Dijk}.} \bibinfo{year}{1996}\natexlab{}.
\newblock \showarticletitle{Forecasting stock market volatility using (non-linear) Garch models}.
\newblock \bibinfo{journal}{\emph{Journal of forecasting}} \bibinfo{volume}{15}, \bibinfo{number}{3} (\bibinfo{year}{1996}), \bibinfo{pages}{229--235}.
\newblock


\bibitem[Gao et~al\mbox{.}(2021)]%
        {gao2021simcse}
\bibfield{author}{\bibinfo{person}{Tianyu Gao}, \bibinfo{person}{Xingcheng Yao}, {and} \bibinfo{person}{Danqi Chen}.} \bibinfo{year}{2021}\natexlab{}.
\newblock \showarticletitle{Simcse: Simple contrastive learning of sentence embeddings}.
\newblock \bibinfo{journal}{\emph{arXiv preprint arXiv:2104.08821}} (\bibinfo{year}{2021}).
\newblock


\bibitem[Gers et~al\mbox{.}(2000)]%
        {gers2000learning}
\bibfield{author}{\bibinfo{person}{Felix~A Gers}, \bibinfo{person}{J{\"u}rgen Schmidhuber}, {and} \bibinfo{person}{Fred Cummins}.} \bibinfo{year}{2000}\natexlab{}.
\newblock \showarticletitle{Learning to forget: Continual prediction with LSTM}.
\newblock \bibinfo{journal}{\emph{Neural computation}} \bibinfo{volume}{12}, \bibinfo{number}{10} (\bibinfo{year}{2000}), \bibinfo{pages}{2451--2471}.
\newblock


\bibitem[Gu et~al\mbox{.}(2020)]%
        {gu2020empirical}
\bibfield{author}{\bibinfo{person}{Shihao Gu}, \bibinfo{person}{Bryan Kelly}, {and} \bibinfo{person}{Dacheng Xiu}.} \bibinfo{year}{2020}\natexlab{}.
\newblock \showarticletitle{Empirical asset pricing via machine learning}.
\newblock \bibinfo{journal}{\emph{The Review of Financial Studies}} \bibinfo{volume}{33}, \bibinfo{number}{5} (\bibinfo{year}{2020}), \bibinfo{pages}{2223--2273}.
\newblock


\bibitem[Khaidem et~al\mbox{.}(2016)]%
        {khaidem2016predicting}
\bibfield{author}{\bibinfo{person}{Luckyson Khaidem}, \bibinfo{person}{Snehanshu Saha}, {and} \bibinfo{person}{Sudeepa~Roy Dey}.} \bibinfo{year}{2016}\natexlab{}.
\newblock \showarticletitle{Predicting the direction of stock market prices using random forest}.
\newblock \bibinfo{journal}{\emph{arXiv preprint arXiv:1605.00003}} (\bibinfo{year}{2016}).
\newblock


\bibitem[Kim and Won(2018)]%
        {kim2018forecasting}
\bibfield{author}{\bibinfo{person}{Ha~Young Kim} {and} \bibinfo{person}{Chang~Hyun Won}.} \bibinfo{year}{2018}\natexlab{}.
\newblock \showarticletitle{Forecasting the volatility of stock price index: A hybrid model integrating LSTM with multiple GARCH-type models}.
\newblock \bibinfo{journal}{\emph{Expert Systems with Applications}}  \bibinfo{volume}{103} (\bibinfo{year}{2018}), \bibinfo{pages}{25--37}.
\newblock


\bibitem[Kruschke(2014)]%
        {kruschke2014doing}
\bibfield{author}{\bibinfo{person}{John Kruschke}.} \bibinfo{year}{2014}\natexlab{}.
\newblock \showarticletitle{Doing Bayesian data analysis: A tutorial with R, JAGS, and Stan}.
\newblock  (\bibinfo{year}{2014}).
\newblock


\bibitem[Lee(2007)]%
        {lee2007application}
\bibfield{author}{\bibinfo{person}{Young-Chan Lee}.} \bibinfo{year}{2007}\natexlab{}.
\newblock \showarticletitle{Application of support vector machines to corporate credit rating prediction}.
\newblock \bibinfo{journal}{\emph{Expert Systems with Applications}} \bibinfo{volume}{33}, \bibinfo{number}{1} (\bibinfo{year}{2007}), \bibinfo{pages}{67--74}.
\newblock


\bibitem[Liapis et~al\mbox{.}(2023)]%
        {liapis2023investigating}
\bibfield{author}{\bibinfo{person}{Charalampos~M Liapis}, \bibinfo{person}{Aikaterini Karanikola}, {and} \bibinfo{person}{Sotiris Kotsiantis}.} \bibinfo{year}{2023}\natexlab{}.
\newblock \showarticletitle{Investigating deep stock market forecasting with sentiment analysis}.
\newblock \bibinfo{journal}{\emph{Entropy}} \bibinfo{volume}{25}, \bibinfo{number}{2} (\bibinfo{year}{2023}), \bibinfo{pages}{219}.
\newblock


\bibitem[Luong et~al\mbox{.}(2015)]%
        {luong2015multi}
\bibfield{author}{\bibinfo{person}{Minh-Thang Luong}, \bibinfo{person}{Quoc~V Le}, \bibinfo{person}{Ilya Sutskever}, \bibinfo{person}{Oriol Vinyals}, {and} \bibinfo{person}{Lukasz Kaiser}.} \bibinfo{year}{2015}\natexlab{}.
\newblock \showarticletitle{Multi-task sequence to sequence learning}.
\newblock \bibinfo{journal}{\emph{arXiv preprint arXiv:1511.06114}} (\bibinfo{year}{2015}).
\newblock


\bibitem[Mohan et~al\mbox{.}(2019)]%
        {mohan2019stock}
\bibfield{author}{\bibinfo{person}{Saloni Mohan}, \bibinfo{person}{Sahitya Mullapudi}, \bibinfo{person}{Sudheer Sammeta}, \bibinfo{person}{Parag Vijayvergia}, {and} \bibinfo{person}{David~C Anastasiu}.} \bibinfo{year}{2019}\natexlab{}.
\newblock \showarticletitle{Stock price prediction using news sentiment analysis}. In \bibinfo{booktitle}{\emph{2019 IEEE fifth international conference on big data computing service and applications (BigDataService)}}. IEEE, \bibinfo{pages}{205--208}.
\newblock


\bibitem[Qin and Yang(2019)]%
        {qin2019you}
\bibfield{author}{\bibinfo{person}{Yu Qin} {and} \bibinfo{person}{Yi Yang}.} \bibinfo{year}{2019}\natexlab{}.
\newblock \showarticletitle{What you say and how you say it matters: Predicting stock volatility using verbal and vocal cues}. In \bibinfo{booktitle}{\emph{Proceedings of the 57th Annual Meeting of the Association for Computational Linguistics}}. \bibinfo{pages}{390--401}.
\newblock


\bibitem[Siami-Namini et~al\mbox{.}(2019)]%
        {siami2019performance}
\bibfield{author}{\bibinfo{person}{Sima Siami-Namini}, \bibinfo{person}{Neda Tavakoli}, {and} \bibinfo{person}{Akbar~Siami Namin}.} \bibinfo{year}{2019}\natexlab{}.
\newblock \showarticletitle{The performance of LSTM and BiLSTM in forecasting time series}. In \bibinfo{booktitle}{\emph{2019 IEEE International conference on big data (Big Data)}}. IEEE, \bibinfo{pages}{3285--3292}.
\newblock


\bibitem[Singh et~al\mbox{.}(2016)]%
        {singh2016review}
\bibfield{author}{\bibinfo{person}{Amanpreet Singh}, \bibinfo{person}{Narina Thakur}, {and} \bibinfo{person}{Aakanksha Sharma}.} \bibinfo{year}{2016}\natexlab{}.
\newblock \showarticletitle{A review of supervised machine learning algorithms}. In \bibinfo{booktitle}{\emph{2016 3rd international conference on computing for sustainable global development (INDIACom)}}. Ieee, \bibinfo{pages}{1310--1315}.
\newblock


\bibitem[Song et~al\mbox{.}(2024)]%
        {song2024omnipred}
\bibfield{author}{\bibinfo{person}{Xingyou Song}, \bibinfo{person}{Oscar Li}, \bibinfo{person}{Chansoo Lee}, \bibinfo{person}{Daiyi Peng}, \bibinfo{person}{Sagi Perel}, \bibinfo{person}{Yutian Chen}, {et~al\mbox{.}}} \bibinfo{year}{2024}\natexlab{}.
\newblock \showarticletitle{Omnipred: Language models as universal regressors}.
\newblock \bibinfo{journal}{\emph{arXiv preprint arXiv:2402.14547}} (\bibinfo{year}{2024}).
\newblock


\bibitem[Souma et~al\mbox{.}(2019)]%
        {souma2019enhanced}
\bibfield{author}{\bibinfo{person}{Wataru Souma}, \bibinfo{person}{Irena Vodenska}, {and} \bibinfo{person}{Hideaki Aoyama}.} \bibinfo{year}{2019}\natexlab{}.
\newblock \showarticletitle{Enhanced news sentiment analysis using deep learning methods}.
\newblock \bibinfo{journal}{\emph{Journal of Computational Social Science}} \bibinfo{volume}{2}, \bibinfo{number}{1} (\bibinfo{year}{2019}), \bibinfo{pages}{33--46}.
\newblock


\bibitem[Vehtari et~al\mbox{.}(2021)]%
        {vehtari2021rank}
\bibfield{author}{\bibinfo{person}{Aki Vehtari}, \bibinfo{person}{Andrew Gelman}, \bibinfo{person}{Daniel Simpson}, \bibinfo{person}{Bob Carpenter}, {and} \bibinfo{person}{Paul-Christian Brkner}.} \bibinfo{year}{2021}\natexlab{}.
\newblock \showarticletitle{Rank-normalization, folding, and localization: An improved R for assessing convergence of MCMC (with discussion)}.
\newblock \bibinfo{journal}{\emph{Bayesian analysis}} \bibinfo{volume}{16}, \bibinfo{number}{2} (\bibinfo{year}{2021}), \bibinfo{pages}{667--718}.
\newblock


\bibitem[Wang et~al\mbox{.}(2023)]%
        {wang2023sparsity}
\bibfield{author}{\bibinfo{person}{Dan Wang}, \bibinfo{person}{Zhi Chen}, \bibinfo{person}{Ionu{\c{t}} Florescu}, {and} \bibinfo{person}{Bingyang Wen}.} \bibinfo{year}{2023}\natexlab{}.
\newblock \showarticletitle{A sparsity algorithm for finding optimal counterfactual explanations: Application to corporate credit rating}.
\newblock \bibinfo{journal}{\emph{Research in International Business and Finance}}  \bibinfo{volume}{64} (\bibinfo{year}{2023}), \bibinfo{pages}{101869}.
\newblock


\bibitem[Wei et~al\mbox{.}(2022)]%
        {wei2022chain}
\bibfield{author}{\bibinfo{person}{Jason Wei}, \bibinfo{person}{Xuezhi Wang}, \bibinfo{person}{Dale Schuurmans}, \bibinfo{person}{Maarten Bosma}, \bibinfo{person}{Fei Xia}, \bibinfo{person}{Ed Chi}, \bibinfo{person}{Quoc~V Le}, \bibinfo{person}{Denny Zhou}, {et~al\mbox{.}}} \bibinfo{year}{2022}\natexlab{}.
\newblock \showarticletitle{Chain-of-thought prompting elicits reasoning in large language models}.
\newblock \bibinfo{journal}{\emph{Advances in neural information processing systems}}  \bibinfo{volume}{35} (\bibinfo{year}{2022}), \bibinfo{pages}{24824--24837}.
\newblock


\bibitem[Xie et~al\mbox{.}(2024)]%
        {xie2024finben}
\bibfield{author}{\bibinfo{person}{Qianqian Xie}, \bibinfo{person}{Weiguang Han}, \bibinfo{person}{Zhengyu Chen}, \bibinfo{person}{Ruoyu Xiang}, \bibinfo{person}{Xiao Zhang}, \bibinfo{person}{Yueru He}, \bibinfo{person}{Mengxi Xiao}, \bibinfo{person}{Dong Li}, \bibinfo{person}{Yongfu Dai}, \bibinfo{person}{Duanyu Feng}, {et~al\mbox{.}}} \bibinfo{year}{2024}\natexlab{}.
\newblock \showarticletitle{The finben: An holistic financial benchmark for large language models}.
\newblock \bibinfo{journal}{\emph{arXiv preprint arXiv:2402.12659}} (\bibinfo{year}{2024}).
\newblock


\bibitem[Yang et~al\mbox{.}(2023)]%
        {yang2023fingpt}
\bibfield{author}{\bibinfo{person}{Hongyang Yang}, \bibinfo{person}{Xiao-Yang Liu}, {and} \bibinfo{person}{Christina~Dan Wang}.} \bibinfo{year}{2023}\natexlab{}.
\newblock \showarticletitle{Fingpt: Open-source financial large language models}.
\newblock \bibinfo{journal}{\emph{arXiv preprint arXiv:2306.06031}} (\bibinfo{year}{2023}).
\newblock


\bibitem[Yang et~al\mbox{.}(2020)]%
        {yang2020html}
\bibfield{author}{\bibinfo{person}{Linyi Yang}, \bibinfo{person}{Tin Lok~James Ng}, \bibinfo{person}{Barry Smyth}, {and} \bibinfo{person}{Riuhai Dong}.} \bibinfo{year}{2020}\natexlab{}.
\newblock \showarticletitle{Html: Hierarchical transformer-based multi-task learning for volatility prediction}. In \bibinfo{booktitle}{\emph{Proceedings of The Web Conference 2020}}. \bibinfo{pages}{441--451}.
\newblock


\bibitem[Yu et~al\mbox{.}(2023)]%
        {yu2023finmem}
\bibfield{author}{\bibinfo{person}{Yangyang Yu}, \bibinfo{person}{Haohang Li}, \bibinfo{person}{Zhi Chen}, \bibinfo{person}{Yuechen Jiang}, \bibinfo{person}{Yang Li}, \bibinfo{person}{Denghui Zhang}, \bibinfo{person}{Rong Liu}, \bibinfo{person}{Jordan~W Suchow}, {and} \bibinfo{person}{Khaldoun Khashanah}.} \bibinfo{year}{2023}\natexlab{}.
\newblock \showarticletitle{FinMem: A performance-enhanced LLM trading agent with layered memory and character design}.
\newblock \bibinfo{journal}{\emph{arXiv preprint arXiv:2311.13743}} (\bibinfo{year}{2023}).
\newblock


\bibitem[Zhang(2020)]%
        {zhang2020financial}
\bibfield{author}{\bibinfo{person}{Beichen Zhang}.} \bibinfo{year}{2020}\natexlab{}.
\newblock \emph{\bibinfo{title}{Financial Risk Disclosure Return Premium: A Topic Modeling Approach}}.
\newblock \bibinfo{thesistype}{Master's\ thesis}. \bibinfo{school}{Stevens Institute of Technology}.
\newblock


\bibitem[Zhang et~al\mbox{.}(2023)]%
        {zhang2023enhancing}
\bibfield{author}{\bibinfo{person}{Boyu Zhang}, \bibinfo{person}{Hongyang Yang}, \bibinfo{person}{Tianyu Zhou}, \bibinfo{person}{Muhammad Ali~Babar}, {and} \bibinfo{person}{Xiao-Yang Liu}.} \bibinfo{year}{2023}\natexlab{}.
\newblock \showarticletitle{Enhancing financial sentiment analysis via retrieval augmented large language models}. In \bibinfo{booktitle}{\emph{Proceedings of the Fourth ACM International Conference on AI in Finance}}. \bibinfo{pages}{349--356}.
\newblock


\bibitem[Zhang et~al\mbox{.}(2024b)]%
        {zhang2024ai}
\bibfield{author}{\bibinfo{person}{Chong Zhang}, \bibinfo{person}{Xinyi Liu}, \bibinfo{person}{Mingyu Jin}, \bibinfo{person}{Zhongmou Zhang}, \bibinfo{person}{Lingyao Li}, \bibinfo{person}{Zhengting Wang}, \bibinfo{person}{Wenyue Hua}, \bibinfo{person}{Dong Shu}, \bibinfo{person}{Suiyuan Zhu}, \bibinfo{person}{Xiaobo Jin}, {et~al\mbox{.}}} \bibinfo{year}{2024}\natexlab{b}.
\newblock \showarticletitle{When AI Meets Finance (StockAgent): Large Language Model-based Stock Trading in Simulated Real-world Environments}.
\newblock \bibinfo{journal}{\emph{arXiv preprint arXiv:2407.18957}} (\bibinfo{year}{2024}).
\newblock


\bibitem[Zhang et~al\mbox{.}(2024a)]%
        {zhang2024benchmarking}
\bibfield{author}{\bibinfo{person}{Tianyi Zhang}, \bibinfo{person}{Faisal Ladhak}, \bibinfo{person}{Esin Durmus}, \bibinfo{person}{Percy Liang}, \bibinfo{person}{Kathleen McKeown}, {and} \bibinfo{person}{Tatsunori~B Hashimoto}.} \bibinfo{year}{2024}\natexlab{a}.
\newblock \showarticletitle{Benchmarking large language models for news summarization}.
\newblock \bibinfo{journal}{\emph{Transactions of the Association for Computational Linguistics}}  \bibinfo{volume}{12} (\bibinfo{year}{2024}), \bibinfo{pages}{39--57}.
\newblock


\bibitem[Zhang et~al\mbox{.}(2024c)]%
        {zhang2024finagent}
\bibfield{author}{\bibinfo{person}{Wentao Zhang}, \bibinfo{person}{Lingxuan Zhao}, \bibinfo{person}{Haochong Xia}, \bibinfo{person}{Shuo Sun}, \bibinfo{person}{Jiaze Sun}, \bibinfo{person}{Molei Qin}, \bibinfo{person}{Xinyi Li}, \bibinfo{person}{Yuqing Zhao}, \bibinfo{person}{Yilei Zhao}, \bibinfo{person}{Xinyu Cai}, {et~al\mbox{.}}} \bibinfo{year}{2024}\natexlab{c}.
\newblock \showarticletitle{FinAgent: A Multimodal Foundation Agent for Financial Trading: Tool-Augmented, Diversified, and Generalist}.
\newblock \bibinfo{journal}{\emph{arXiv preprint arXiv:2402.18485}} (\bibinfo{year}{2024}).
\newblock


\end{thebibliography}

\appendix

\begin{figure*}[ht]
    \centering
{\includegraphics[width=0.4\textwidth]{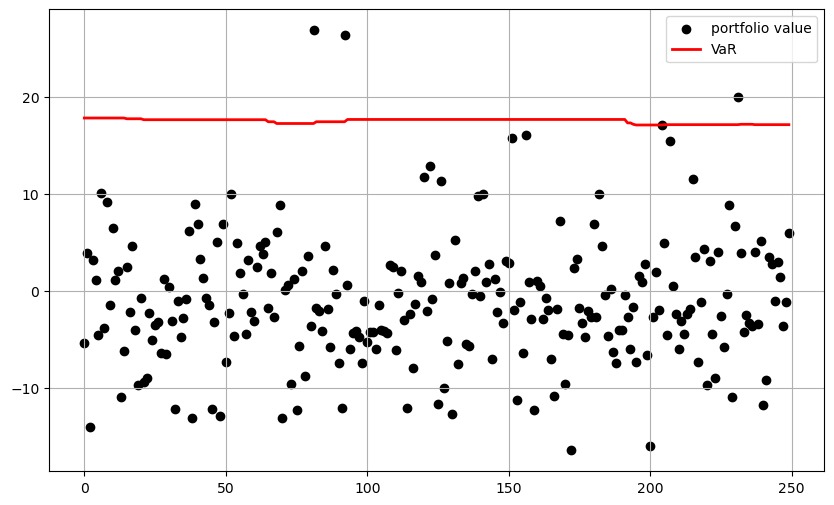}\label{fig:a}}
\hfil
{\includegraphics[width=0.4\textwidth]{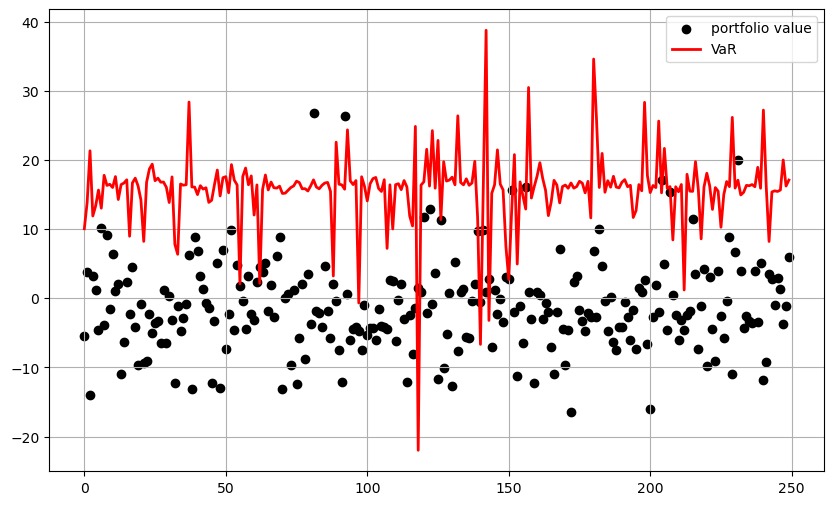}\label{fig:b}}
\caption{This figure presents two plots that compare daily Value at Risk predictions (red curve) with the actual returns of the asset (black dot). 
It visualizes the percentage of actual returns exceeding the predicted VaR. For instance, with a predefined VaR of 0.05, observing approximately 5\% of the actual returns surpassing the predicted VaR curve indicates a high degree of prediction accuracy.
The left plot showcases the VaR forecast using the historical method, illustrating how this traditional technique estimates risk in relation to the asset's actual performance. The right plot, on the other hand, using a fully connected neural network for VaR prediction, offering a modern computational approach to risk assessment.}
\label{fig: Plot of residuals against the predicted values}
\end{figure*}

\section{Experiment Details}
\label{app:exp}
\subsection{Dataset} 
\label{app:dataset}
For dataset usage, we strategically partitioned it into a training set and a test set, adhering to an 8:2 ratio. It is crucial to highlight that the division of the data was conducted on a temporal basis, ensuring that the dates of the data in the training set always precede those in the test set. This temporal segregation is imperative for maintaining the integrity of our predictive model. By structuring the dataset in this manner, we ensure that the training process is consistently oriented towards predicting future risks based on past data, a fundamental principle for the accuracy and reliability of our forecasting methodology.
\subsection{Baseline Setup}
\label{app:baseline}
We compare our approach to volatility prediction to several important baselines as described below.
\begin{itemize}
    \item \textbf{Classical Methods:} The analysis incorporates the GARCH model, a classical auto-regressive model for predicting volatility, as well as its various derivatives~\cite{franses1996forecasting, kim2018forecasting}. These models are widely recognized and commonly employed in the realm of volatility prediction. Primarily, they are tailored for short-term volatility forecasting and may not perform as efficiently in predicting average volatility over an extended period (e.g., n-day volatility). 
    \item \textbf{LSTM~\cite{gers2000learning}}: Long Short-Term Memory Networks (LSTMs) are a popular choice for financial time series prediction due to their efficacy in handling sequential data. In the context of volatility prediction, we select a straightforward LSTM model as a benchmark.
    \item \textbf{MT-LSTM+ATT~\cite{luong2015multi}:} merges the prediction of average n-day volatility with the forecasting of single-day volatility, using attention-enhanced LSTM as the foundational learning models.
    \item \textbf{HAN (Glove):} This baseline implements a Hierarchical Attention Network that incorporates dual-layered attention mechanisms at both the word and sentence levels. Initially, every word within a sentence is transformed into a word embedding through the pre-trained Glove 300-dimensional embeddings. Subsequently, these embedded sentences are processed by a Bi-GRU encoder~\cite{chung2014empirical}, and in parallel, another Bi-GRU encoder is utilized to formulate a representation of each document as a series of sentences. This representation of the document is then fed into the concluding regression layer to generate predictions.
    \item \textbf{MRDM~\cite{qin2019you}:} The MRDM model first introduced a multi-modal deep regression approach for volatility prediction tasks. It utilizes pre-trained GloVe embeddings and bespoke acoustic features, which are processed through individual BiLSTMs to generate uni-modal contextual embeddings. These embeddings are subsequently merged and input into a two-layer dense network for further processing.
    \item \textbf{HTML~\cite{yang2020html}:} This work presented a state-of-the-art model that employs WWM-BERT for text token encoding. Similar to MDRM, HTML also leverages the same audio features. These unimodal features are then combined and processed through a sentence-level transformer, resulting in multimodal representations for each call.
    \item \textbf{GPT-3.5-Turbo:} We evaluated the efficacy of using GPT-3.5-Turbo for direct financial risk prediction. The input for this test was earnings conference calls, and we instructed GPT-3.5-Turbo to generate numerical risk forecasts based on the provided EC, utilizing a specified prompt setting. In the experiment, we set the temperature as zero.
\end{itemize}

\section{Detailed Experiment Results}
\label{app:result}
\subsection{Detailed analysis of VaR values}
\label{app:var}

From the analysis presented in the figure, a notable observation emerges: the plot on the left, which uses the historical method for VaR prediction, appears relatively flat, indicating a consistent, albeit less responsive, forecast over time. In contrast, the plot on the right, using a fully connected neural network, exhibits a more zigzag pattern, reflecting greater responsiveness to daily information changes. This contrast suggests that AI techniques, such as neural networks or LLMs, offer a dynamic advantage by more effectively incorporating daily updates into the model, as opposed to relying solely on historical scenarios. The ability of AI-driven models to adapt to new information underscores their potential for providing more accurate and timely risk assessments.

\end{document}